\documentclass[sigconf]{acmart}
\usepackage[normalem]{ulem}
\useunder{\uline}{\ul}{}
\usepackage{graphicx}
\usepackage{subcaption}
\usepackage{amsmath}
\usepackage{enumitem}
\usepackage{balance}
\AtBeginDocument{%
  }

\setcopyright{acmlicensed}
\copyrightyear{2025}
\acmYear{2025}
\acmDOI{10.1145/3701716.3715589}

\acmConference[WWW Companion '25]{Companion Proceedings of the ACM Web Conference 2025}{April 28-May 2, 2025}{Sydney, NSW, Australia}
\acmISBN{979-8-4007-1331-6/25/04}
\acmDOI{10.1145/3701716.3715589}

\begin{document}

\title{PCL: Prompt-based Continual Learning for User Modeling in Recommender Systems}


\author{Mingdai~Yang}
\email{myang72@uic.edu}
\affiliation{%
  \institution{University of Illinois at Chicago}
  \city{Chicago}
  \country{USA}}
\author{Fan~Yang}
\email{fnam@amazon.com}
\affiliation{%
  \institution{Amazon}
  \city{Seattle}
  \country{USA}}
\author{Yanhui~Guo}
\email{yanhuig@amazon.com}
\affiliation{%
  \institution{Amazon}
  \city{Seattle}
  \country{USA}}
\author{Shaoyuan~Xu}
\email{shaoyux@amazon.com}
\affiliation{%
  \institution{Amazon}
  \city{Seattle}
  \country{USA}}
  
\author{Tianchen~Zhou}
\email{tiancz@amazon.com}
\affiliation{%
  \institution{Amazon}
  \city{Seattle}
  \country{USA}} 
\author{Yetian~Chen}
\email{yetichen@amazon.com}
\affiliation{%
  \institution{Amazon}
  \city{Seattle}
  \country{USA}} 
\author{Simone~Shao}
\email{simengsh@amazon.com}
\affiliation{%
  \institution{Amazon}
  \city{Seattle}
  \country{USA}} 

\author{Jia Liu}
\email{liu@ece.osu.edu}
\affiliation{%
   \institution{Amazon~\& The Ohio State University}
   \country{Columbus, USA}}
   
\author{Yan~Gao}
\email{yanngao@amazon.com}
\affiliation{%
  \institution{Amazon}
  \city{Seattle}
  \country{USA}}
  
\renewcommand{\shortauthors}{Mingdai Yang et al.}

\begin{abstract}
User modeling in large e-commerce platforms aims to optimize user experiences by incorporating various customer activities. Traditional models targeting a single task often focus on specific business metrics, neglecting the comprehensive user behavior, and thus limiting their effectiveness. To develop more generalized user representations, some existing work adopts Multi-task Learning (MTL) approaches. But they all face the challenges of optimization imbalance and inefficiency in adapting to new tasks. Continual Learning (CL), which allows models to learn new tasks incrementally and independently, has emerged as a solution to MTL's limitations. However, CL faces the challenge of catastrophic forgetting, where previously learned knowledge is lost when the model is learning the new task. Inspired by the success of prompt tuning in Pretrained Language Models (PLMs), we propose PCL, a \textbf{P}rompt-based  \textbf{C}ontinual  \textbf{L}earning framework for user modeling, which utilizes position-wise prompts as external memory for each task, preserving knowledge and mitigating catastrophic forgetting. Additionally, we design contextual prompts to capture and leverage inter-task relationships during prompt tuning. We conduct extensive experiments on real-world datasets to demonstrate PCL's effectiveness.
\end{abstract}

\ccsdesc[500]{Information systems~Retrieval models and ranking}

\keywords{Prompt Tuning; Continual Learning; User Modeling}


\maketitle

\section{Introduction}
User modeling has been widely applied in large e-commerce platforms to interpret and optimize user experience. 
Most machine learning models for user modeling optimize for their respective business metrics without jointly considering all activities performed by customers\cite{Jiao2024,sasrec18}, resulting in suboptimal performance. 
For instance, a sequential recommendation model trained only on users' purchase sequences could significantly benefit from incorporating auxiliary relations and multiple behaviors from users and items (e.g., users' delivery preferences, co-clicks, and return behaviors) to optimize its tasks \cite{liu2025,yang2022multi,zhang2019feature}. 
Thus, for a seamless and consistent customer experience throughout the shopping journey, it is highly desirable to build a universal representation that incorporates all data modes to easily adapt to a large variety of applications and scale across multiple marketplaces at an affordable cost. 
To date, most existing works adopt multi-task learning (MTL) to learn a generalized representation by simultaneously training a single model on multiple tasks~\cite{mmoe18,yabo18,UPRTH}. 
However, MTL methods often suffer from optimization imbalance~\cite{mgda18,10.5555/3692070.3694632}.
What's worse is that, when training on a new incoming task, MTL requires full model retraining together with all previous tasks, which is inefficient or even infeasible.

To address the challenges of sequentially arriving new tasks under MTL, Continual Learning (CL) has recently been introduced~\cite{teracon23,conure21}, 
which aims to learn each new task incrementally without retraining from previous tasks.
However, in CL, a major challenge is to mitigate the catastrophic forgetting (CF) problem, i.e., the learned knowledge from previous tasks is lost when the model is learning a new task. 
CONURE~\cite{conure21} addresses CF by training part of the model parameters for each task, but this restricts the model's learning capability due to the reduction of available per-task learnable parameters.
TERACON~\cite{teracon23} retains learned knowledge by regularizing the model updates between consecutive tasks. However, it requires continually updating the backbone and the adapters of all previous tasks, which is computationally inefficient.

With the success of prompt tuning in Pretrained Language Models (PLMs), prompt tuning has been leveraged in CL for mitigating CF of sequential text-based and image-based classification tasks~\cite{qtune24,liyuan23}. 
Prompt tuning updates a pretrained model with low memory and computational costs, since only a small number of parameters corresponding to the prompt of each task are finetuned~\cite{YuZ0024,YangLYLWPY24}. 
However, so far, how to apply prompt tuning for continual user modeling in the recommender systems design remains under-explored. 
Notably, tasks in user modeling, such as user age prediction and next thumbs-up item prediction, are more heterogeneous than classifying text or images with different groups of labels. 
Many features in each task are unique and could vary dramatically, such as user behaviors, user profiles, and item categories. 
In this work, we propose a Prompt-based Continual Learning (PCL) framework for continual user modeling tasks in recommender systems, which leverages prompts to enable task adaptation and knowledge retention among consecutive tasks. 
We design position-wise prompts in the form of external memory for each individual task. 
The position-wise prompt for each task inherits and retains the knowledge from previously trained tasks during new task adaptation. 
To guide prompt tuning with inter-task contextual relations, we propose a new contextual prompt-based on the attention between tasks learned during downstream finetuning. 
Our main contributions to this work are summarized as follows:

\begin{itemize}[left=0pt]
    \item We propose a new PCL framework that leverages prompts for continual user modeling in recommender systems design. To our knowledge, PCL is the first work to mitigate catastrophic forgetting by prompt tuning in continual user modeling.
    \item To capture and exploit relationships between tasks during prompt tuning, we design simple and effective contextual prompts based on the attention mechanism.
    \item We conduct extensive experiments on two real-world datasets to i) validate the efficacy of PCL and ii) demonstrate its robustness to different task orders and cold-start settings.
\end{itemize}




\section{Problem Formulation}
We use $\mathcal{T}=\{T_1,T_2,...,T_{|\mathcal{T}|}\}$ to denote the set of sequentially arriving tasks.
We let $\mathcal{I}$ and $\mathcal{U}$ represent the item set and the user set, respectively. 
In each task, every user $u\in \mathcal{U}$ is associated with a behavior sequence $\mathbf{s}^u=\{i^u_1, i^u_2,..., i^u_n\}$, where $i^u_j\in \mathcal{I}$ is the $j$-th item interacted by the user $u$. Each task $T_k$ is associated with a set of labels $\mathbf{Y}^k$ with $|\mathbf{Y}^k| = |\mathcal{U}|$. In other words, each user has a label to be predicted in each task, such as the category of the next clicked item in next-category prediction. 
Our goal is to train a single model for all tasks in a sequential manner to predict the label of each user in each task. 
Upon the completion of training on all tasks, this model will be applied in the inference stage without further training. In this work, next item prediction is always used as the first task $T_1$ due to its self-supervised nature. 
A recommendation backbone model is pretrained on $T_1$ to capture and preserve user-item interaction information before prompt tuning for other downstream tasks.

\section{The Proposed PCL Framework}
\begin{figure*}[h]
  \centering
    \includegraphics[width=\linewidth]{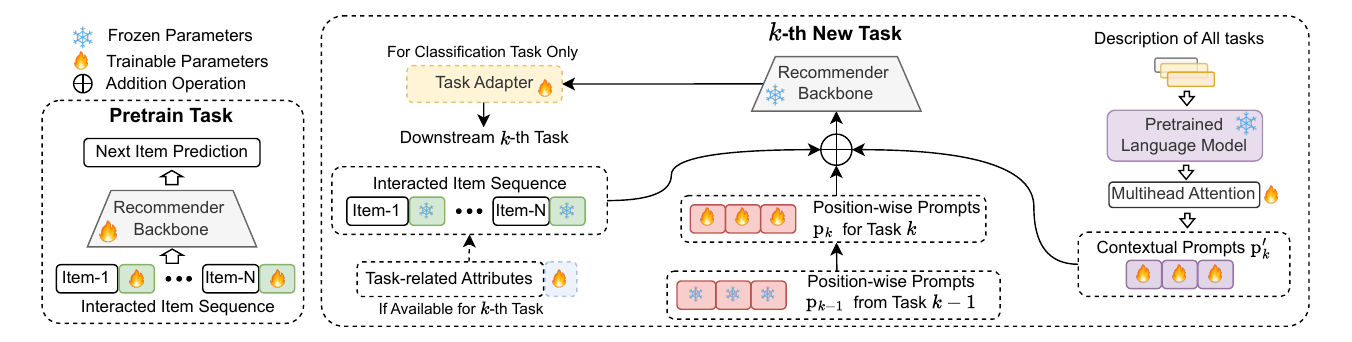}
    
  \caption{The overall framework of PCL. The backbone model and item embeddings are frozen in the downstream tasks.}
  \label{fig:PCL}
  
\end{figure*}

\subsection{Position-wise Prompts}
With a frozen backbone model and its item embeddings pretrained in a self-supervised fashion on the next-click prediction task $T_1$, we expand the model's capability by introducing a prompt for each of the downstream tasks. 
Formally, let $\mathbf{p}_k\in\mathbb{R}^{n\times d}$ denote the learnable position-wise prompt sequence for task $T_k$, where $n$ is the maximum length of behavior sequences and $d$ is the dimension of the pretrained item embedding. 
In user modeling based on behavior sequences, items interacted by a user more recently reflect more recent preferences and features for this user. Thus, for each behavior sequence, we only fuse the last $t$ item embeddings with their corresponding position-wise prompts. Assuming the backbone model has been pretrained by a behavior sequence $\mathbf{s}^u$ in self-supervised task $T_1$, we prompt the behavior embedding sequence as follows:
\begin{equation}\label{eq:initial}
    \mathbf{E}^{\mathbf{s}^u}_k = \mathbf{E}^{\mathbf{s}^u}_1[:-t,:] \|(\mathbf{E}^{\mathbf{s}^u}_1[-t:,:]\oplus\mathbf{p}_k[-t:,:]),
\end{equation}
where $\mathbf{E}^{\mathbf{s}^u}_1[-t:,:]$ represents the embeddings of the last $t$ items in the behavior sequence and $\mathbf{E}^{\mathbf{s}^u}_1[:-t,:]$ denotes the embeddings of the remaining items in the sequence. 
We fuse the last $t$ prompts of the position-wise prompt sequence $\mathbf{p}_k[-t:,:]$ with the last $t$ item embeddings by element-wise addition $\oplus$, and then concatenate the fused embedding with the frozen first $n-t$ item embeddings.
By doing so, the most relevant prior knowledge for the task $k$ is extracted by prompt tuning.


The total memory cost of prompts for all tasks is $|\mathcal{T}|\times n\times d$, which is negligible in comparison to the number of backbone model parameters and the item embedding table size $|\mathcal{I}|\times d$.

\subsection{Contextual Prompts}
In addition to the position-wise prompts that allow a frozen model for task adaptation by expanding learning capacity, we further introduce contextual prompts to guide the prompt tuning by both task information and inter-task relations. 
A pretrained sentence transformer~\cite{minilm} is deployed to encode task descriptions as text embeddings $\mathbf{E_{\mathcal{T}}}\in \mathbb{R}^{|\mathcal{T}|\times d}$, which are frozen in all downstream tasks. Each task description consists of the task name, the input data format, the output data format and the evaluation metric. A multi-head attention layer~\cite{transformer} is used to adaptively learn the relations among all the tasks:
\begin{equation}\label{eq:ta}
\begin{aligned}
\mathbf{P}_\mathcal{T} &= \text{Concat}(\text{head}_1, ..., \text{head}_h)W^O,\\
\text{head}_m &= \text{Attention}(\mathbf{E_{\mathcal{T}}}W^Q_m,\mathbf{E_{\mathcal{T}}}W^K_m,\mathbf{E_{\mathcal{T}}}W^V_m),
\end{aligned}
\end{equation}
where $W^Q_m,W^K_m,W^V_m\in\mathbb{R}^{d\times d_h}$ and $W^O\in\mathbb{R}^{d\times d}$ are projection matrices. 
We obtain the contextual prompts $\mathbf{P}_\mathcal{T} = [\mathbf{p}'_1,...,\mathbf{p}'_{|\mathcal{T}|}]$ for all tasks in $\mathcal{T}$. Similar to the fusion between pretrained item embeddings and position-wise prompts, we add the contextual prompts to each of the last $t$ embeddings in the behavior sequence:
\begin{equation}\label{eq:task}
    \mathbf{E'}^{\mathbf{s}^u}_k = \mathbf{E}^{\mathbf{s}^u}_k[:-t,:] \|(\mathbf{E}^{\mathbf{s}^u}_k[-t:,:]+\lambda\mathbf{p}'_k),
\end{equation}
where $\lambda$ denotes the prompting intensity parameter. The effectiveness of the proposed contextual prompts is verified in Section~\ref{sec:contextual prompt}.
\subsection{Pretraining and Prompt Tuning}

Our PCL framework is illustrated in Figure~\ref{fig:PCL}. 
To capture temporal dynamics in user behavior, we first pretrain a sequential recommendation backbone model $\mathcal{M}$ with all item embeddings $\mathbf{N}\in\mathbb{R}^{|\mathcal{I}|\times d}$ on the self-supervised next-item prediction task $T_1$. For link prediction tasks, we adopt the binary cross entropy loss for sequence recommendation~\cite{sasrec18} as the objective function.
For attribute classification tasks, the cross entropy loss $\mathcal{L}_{CE}(f_k(\mathcal{M}(\mathbf{E'}^{\mathbf{s}^u}_k)),\mathbf{y}^k_u)$ is adopted as the objective function,
where $\mathbf{y}^k_u$ is the ground-truth label, $f_k(\cdot)$ represents the task adapter, which can be a linear layer or an MLP model to map the dimension of latent vectors to the number of classes for classification.

During the prompt tuning stage, the position-wise prompts are randomly initialized in $T_2$. From $T_3$ to $T_{|\mathcal{T}|}$, the prompts $\mathbf{p}_k$ are initialized as $\mathbf{p}_{k-1}$, which is the optimized prompt for the previous task. 
In addition, our framework is designed to leverage task-related attributes when they are available in downstream tasks. This capability allows the model to incorporate additional contextual information, enhancing its adaptability to specific task requirements. 
For example, in the classification of the next item's category in a behavior sequence, a feature vector sequence containing each item's category embedding is added to and finetuned with the frozen item sequence.
In this work, we use SASRec~\cite{sasrec18} as the backbone model and the Adam~\cite{adam14} method as the optimizer.

\section{Experiment}
\subsection{Experiment Setup}
\textbf{1) Dataset:} We conduct experiments on two real-world datasets: Tenrec~\cite{tenrec22} and MovieLens~\footnote{https://grouplens.org/datasets/movielens/25m/}. 
We sample the datasets with fewer user-item interactions to provide a more challenging evaluation environment. Tenrec involves two link prediction tasks and two attribute classification tasks on 15,046 users and 16,475 items from the Kandian platform. 
For Tenrece, $T_1$ contains the users’ recent 50 news $\&$ video-watching interactions, $T_2$ aims to predict next thumb-up items for users, $T_3$ aims to classify categories of next-click items, and $T_4$ is to predict users' ages. 
MovieLens contains three link prediction tasks and one classification task on 20,741 users and 10,033 items from the MovieLens website. 
Specifically, $T_1$ contains users' recent 50 clicking interactions, $T_2$ and $T_3$ aim to predict next four-starred and five-starred movies rated by users, and $T_4$ is trained to classify genres of the next clicked movies. We select Hit Ratio (HR@5 and HR@10) and accuracy (Acc) as the metrics for link prediction and attribute classification, respectively. 
We use the batch-ranking~\cite{sasrec18} and all-ranking protocols in MovieLens and Tenrec, respectively, for a comprehensive HR evaluation.

\noindent \textbf{2) Baselines:} To show the effectiveness of PCL, we compare it with four single-objective baselines. 
\textit{a) SASRec} trains a single model for each task from scratch without knowledge transfer between tasks. 
\textit{b) SinMo} trains a single model for all tasks without saving parameters from previous tasks.
\textit{c) FineAll} finetunes all the model for each task after training on $T_1$ for transfer learning.
\textit{d) Teracon}~\cite{teracon23} is a CL-based method leveraging soft masks and function regularization for knowledge transfer.
We also compare PCL with four MTL baselines that optimize all tasks simultaneously.
\textit{MTL}~\cite{rich98} is a standard multi-task optimization method via parameter sharing.
\textit{MMOE}~\cite{mmoe18} is an MTL method based on mixture-of-experts.
\textit{MTL+} and \textit{MMOE+} apply a multi-gradient descent algorithm~\cite{mgda18} to dynamically determine the weight of each objective to address gradient conflicts.
In addition to overall performance comparison, we design experiments to answer the following three research questions: 
\begin{itemize}
    \item \textbf{RQ1:} Does the proposed contextual prompts benefit PCL?
    \item \textbf{RQ2:} Is the performance of PCL robust to the order of tasks?
    \item \textbf{RQ3:} Does PCL improve downstream finetuning with cold-start items?
    \item \textbf{RQ4:} Can PCL generate universal user representations for unseen models and tasks?
\end{itemize}

\begin{table*}[]\caption{Overall performance comparison. $\#$B is the number of backbone models to be preserved for inference on all tasks. $\#$F is the number of objective tasks to be optimized simultaneously. Multi-task learning methods simultaneously supervised by all tasks with $\#$F$=4$ are compared as the upper bounds of CL baselines.}\label{tab:overall}

\resizebox{0.95\textwidth}{!}{
\begin{tabular}{l|cccccc|ccccclc|cc}
\hline
        & \multicolumn{6}{c|}{Tenrec}                                                                                                                                                                                                                                                                                                                & \multicolumn{7}{c|}{MovieLens}                                                                                                                                                                                                                                                                                                                                                &     &     \\ \cline{2-14}
Method  & \multicolumn{2}{c}{$T_1$}                                                                                           & \multicolumn{2}{c}{$T_2$}                                                                             & $T_3$                                             & $T_4$                                                    & \multicolumn{2}{c}{$T_1$}                                                                                    & \multicolumn{2}{c}{$T_2$}                                                                                           & \multicolumn{2}{c}{$T_3$}                                                            & $T_4$                                             & \#B & \#F \\
        & HR@5                                                     & HR@10                                                    & HR@5                                              & HR@10                                             & ACC                                               & ACC                                                      & HR@5                                                     & HR@10                                             & HR@5                                                     & HR@10                                                    & HR@5                                                     & \multicolumn{1}{c}{HR@10} & ACC                                               &     &     \\ \hline
SASRec  & {\ul 0.6582}                                                   & {\ul 0.8112}                                                  & 0.6613                                            & 0.8107                                            & 0.3329                                            & 0.7722                                                   & {\ul 0.3018}                                             & {\ul 0.4450}                                      & 0.3007                                                   & 0.4627                                                   & 0.2859                                                   & 0.4369                    & {\ul 0.4783}                                      & 4   & 1   \\ \hline
SinMo   & 0.1028                                                   & 0.1777                                                   & 0.1083                                            & 0.1874                                            & 0.0581                                            & 0.7553                                                   & 0.1157                                                   & 0.1876                                            & 0.0937                                                   & 0.1651                                                   & 0.0890                                                   & 0.1558                    & 0.4696                                            & 1   & 1   \\
FineAll & -                                                        & -                                                        & {\ul 0.6832}                                      & {\ul 0.8292}                                      & {\ul 0.3409}                                      & {\ul 0.7837}                                             & -                                                        & -                                                 & {\ul 0.3291}                                             & {\ul 0.4785}                                             & 0.3015                                                   & 0.4547                    & \textbf{0.4812}                                   & 4   & 1   \\
Teracon & 0.5703                                             &  0.7388                                             & 0.6390                                            & 0.7804                                            & 0.3071                                            & 0.7725                                                   & 0.1177                                                   & 0.1704                                            & 0.2878                                                   & 0.3986                                                   & \textbf{0.4297}                                          & \textbf{0.5594}           & 0.3540                                            & 1   & 1   \\ \hline
PCL     & \textbf{0.6582}                                          & \textbf{0.8112}                                          & \textbf{0.6936}                                   & \textbf{0.8372}                                   & \textbf{0.3466}                                   & \textbf{0.7974}                                          & \textbf{0.3018}                                          & \textbf{0.4450}                                   & \textbf{0.3315}                                          & \textbf{0.4845}                                          & {\ul 0.3155}                                             & {\ul 0.4615}              & 0.4634                                            & 1   & 1   \\ \hline
MTL     & 0.6361                                                   & 0.7965                                                   & 0.6968                                            & 0.8355                                            & 0.3285                                            & 0.7447\textsuperscript{$\triangledown$} & 0.2816\textsuperscript{$\triangledown$} & 0.4364                                            & 0.3036                                                   & 0.4603                                                   & 0.3002                                                   & 0.4414                    & 0.4744                                            & 1   & 4   \\
MMOE    & 0.6565\textsuperscript{$\triangledown$} & 0.8085\textsuperscript{$\triangledown$} & 0.6884                                            & 0.8351                                            & 0.3227                                            & 0.7385                                                   & 0.2517                                                   & 0.3631                                            & 0.2521                                                   & 0.3734                                                   & 0.2460                                                   & 0.3682                    & 0.4609                                            & 1   & 4   \\
MTL+    & 0.6489                                                   & 0.8045                                                   & 0.7043\textsuperscript{$\Delta$} & 0.8434\textsuperscript{$\Delta$} & 0.3528                                            & 0.7434                                                   & 0.2811                                                   & 0.4503\textsuperscript{$\Delta$} & 0.3106\textsuperscript{$\triangledown$} & 0.4707\textsuperscript{$\triangledown$} & 0.3020\textsuperscript{$\triangledown$} & 0.4522\textsuperscript{$\triangledown$}                  & 0.4759\textsuperscript{$\Delta$} & 1   & 4   \\
MMOE+   & 0.6157                                                   & 0.7899                                                   & 0.6823                                            & 0.8331                                            & 0.3551\textsuperscript{$\Delta$} & 0.6166                                                   & 0.2536                                                   & 0.3732                                            & 0.2659                                                   & 0.3952                                                   & 0.2481                                                   & 0.3725                    & 0.4716                                            & 1   & 4   \\ \hline
\end{tabular}
}

\end{table*}
\subsection{Overall Performance}
We show the results in Table~\ref{tab:overall}. 
The best and second-best single-objective methods are in boldface and underlined. 
The best MTL method better/worse than PCL is marked with $\textsuperscript{$\triangle$}$/$\textsuperscript{$\triangledown$}$. 
We have the following observations: 1) With only one copy of backmode parameters preserved, PCL achieves the best in 6 out of 8 tasks compared with other single-objective baselines. This justifies the advantages of prompt tuning for continual user modeling; and 2) Even if the supervised signals from other tasks are not available when training on the target task, PCL is generally better than all MTL baselines that leverage labels in all tasks for optimization. 
This verifies the efficacy of prompts as external memory for task adaptation.

\subsection{RQ1: Impact of Contextual Prompts}\label{sec:contextual prompt}

\begin{table}[H]\caption{Ablation study of contextual prompts.}\label{tab:contextual prompts}

\resizebox{0.45\textwidth}{!}{
\begin{tabular}{l|ccc|ccc}
\hline
        & \multicolumn{3}{c|}{Tenrec} & \multicolumn{3}{c}{MovieLens} \\ \cline{2-7} 
Method  & $T_2$ & $T_3$ & $T_4$ & $T_2$ & $T_3$ & $T_4$ \\
        & HR@5 & ACC & ACC & HR@5 & HR@5 & ACC \\ \hline
w/o $\mathbf{P}_\mathcal{T}$ & 0.6927 & 0.3404 & 0.7908 & 0.3259 & 0.3083 & 0.4431 \\
w/o PLM & 0.6871 & 0.3347 & 0.7886 & 0.3269 & 0.3055 & 0.4566 \\ \hline
PCL     & \textbf{0.6936} & \textbf{0.3449} & \textbf{0.7974} & \textbf{0.3315} & \textbf{0.3155} & \textbf{0.4634} \\ \hline
\end{tabular}
}

\end{table}
To quantify the contribution of contextual prompts in our framework, we compare the performance of PCL to its variants (i) without contextual prompts $\mathbf{P}_{\mathcal{T}}$ and (ii) without PLM to initialize contextual prompts from task descriptions. For the variant of PCL without PLM, we randomly initialize the contextual prompts. The results are shown in Table~\ref{tab:contextual prompts}. We find that employing PLM to encode task descriptions leads to stable enhancement in both datasets. This implies contextual prompts in PCL effectively capture and leverage task relations during sequential task learning.

\subsection{RQ2: Robustness against Task Orders}
\begin{figure}
    \begin{subfigure}{0.15\textwidth}
    \includegraphics[width=\textwidth]{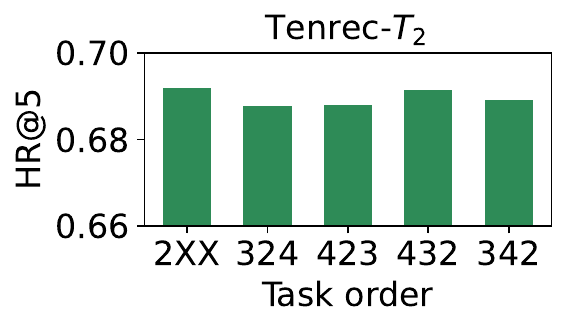}
    \end{subfigure}
    \hfill
    \begin{subfigure}{0.15\textwidth}
    \includegraphics[width=\textwidth]{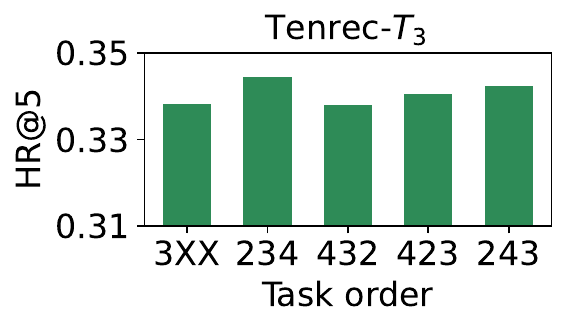}
    \end{subfigure}
    \hfill
    \begin{subfigure}{0.15\textwidth}
    \includegraphics[width=\textwidth]{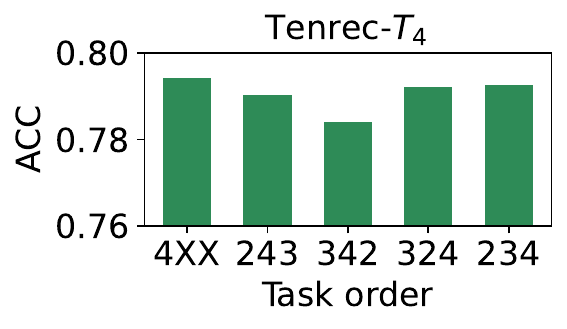}
    \end{subfigure}

    \begin{subfigure}{0.15\textwidth}
    \includegraphics[width=\textwidth]{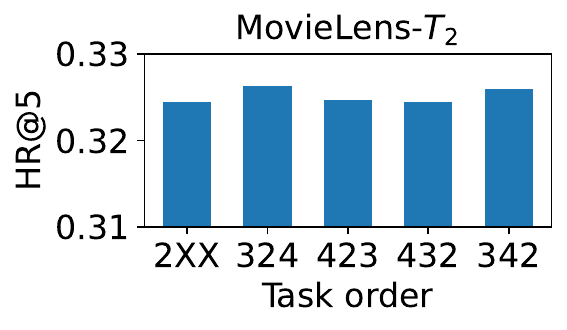}
    \end{subfigure}
    \hfill
    \begin{subfigure}{0.15\textwidth}
    \includegraphics[width=\textwidth]{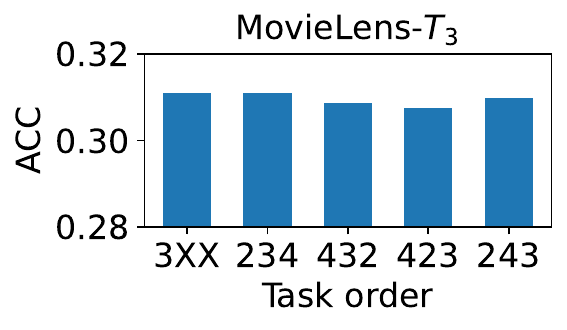}
    \end{subfigure}
    \hfill
    \begin{subfigure}{0.15\textwidth}
    \includegraphics[width=\textwidth]{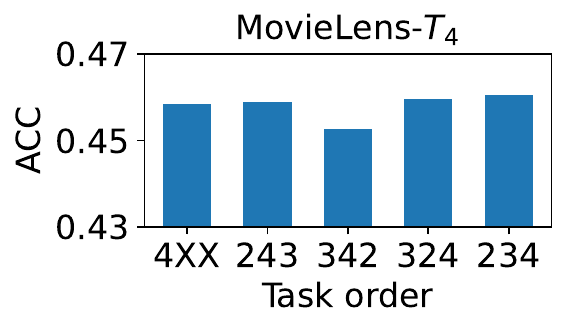}
    \end{subfigure}
    
        \caption{Performance of each task w.r.t the task order.}~\label{fig:task orders}
\end{figure}

Considering that the sequence of tasks cannot be arbitrarily determined in real-world scenarios, we verify the robustness of PCL against task orders by conducting experiments on all permutations of downstream tasks. The results are shown in Figure~\ref{fig:task orders}, where each coordinate on the x-axis represents a task order. For example, \textit{4XX} means the performance of $T_4$ when $T_4$ is the first downstream task after pretraining on the self-supervised $T_1$. We observe that the average performance degradation from the best to the worst task order on all tasks is only $1.20\%$ in both datasets. Since each downstream task is trained with independent prompts, the performance of each task is robust to the training order.

\subsection{RQ3: Finetuning with Cold-start Items}
\begin{figure}

    \begin{subfigure}{0.153\textwidth}
    \includegraphics[width=\textwidth]{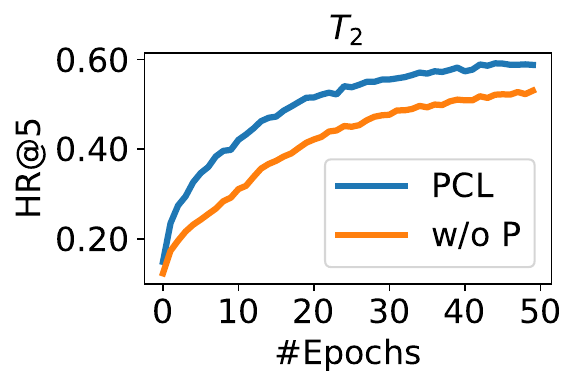}
    \end{subfigure}
    \hfill
    \begin{subfigure}{0.153\textwidth}
    \includegraphics[width=\textwidth]{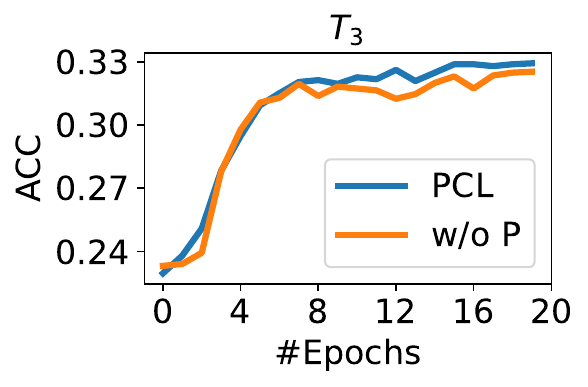}
    \end{subfigure}
    \hfill
    \begin{subfigure}{0.153\textwidth}
    \includegraphics[width=\textwidth]{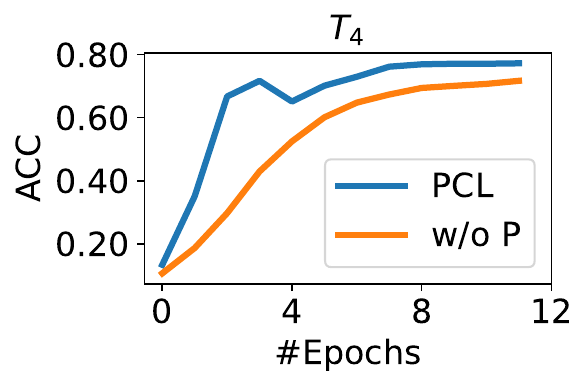}
    \end{subfigure}
    
        \caption{Performance w.r.t. the number of training epochs with $50\%$ cold-start items in Tenrec dataset when $lr=0.01$.}\label{fig:cold start}
\end{figure}

We investigate the efficacy of PCL in finetuning with cold-start items by masking $50\%$ items in $T_1$ as cold-start items. The performance in the following tasks is shown as curves in Figure~\ref{fig:cold start}, compared with only finetuning cold-start item embeddings and task adapters without prompts. Benefiting from the adaptability of prompt tuning, PCL converges to better performance without extensive retraining. 
The necessity of prompt tuning stands out more in $T_2$ without available task adapters.

\subsection{RQ4: Universal User Representation}
\begin{figure}
    \begin{subfigure}{0.20\textwidth}
    \includegraphics[width=\textwidth]{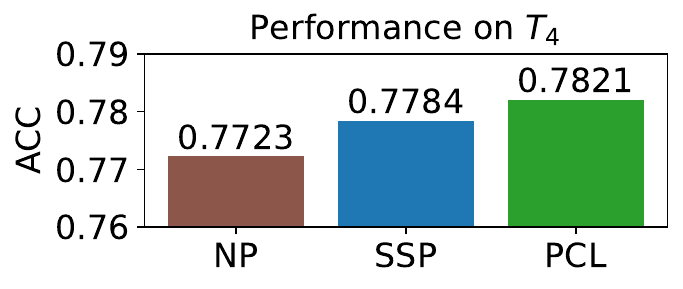}
    \end{subfigure}
    \hspace{1mm}
    \begin{subfigure}{0.19\textwidth}
    \includegraphics[width=\textwidth]{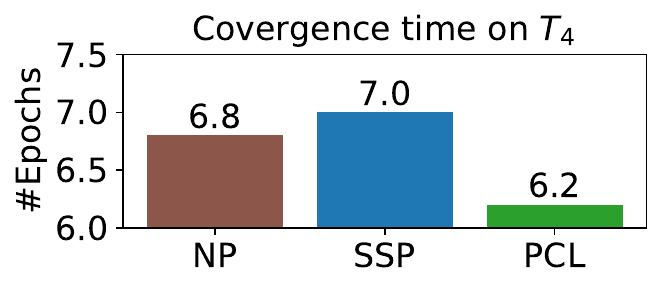}
    \end{subfigure}
        \caption{Performance and convergence time of an MLP model on $T_4$ in Tenrec with user features obtained from different pretraining methods.}\label{fig:case study}
\end{figure}
We further conduct a case study to verify that PCL can generate a universal user representation through prompt tuning. Besides the downstream tasks where the prompts are finetuned, this representation can be leveraged as an informative user feature in new tasks with a model other than the backbone. To be concrete, we finetune the prompts on $T_2$ and $T_3$ in Tenrec, and the behavior latent vector output from the model is used as the universal representation of each user. We concatenate these universal representations as fixed user features to one-hot user vectors and feed them into a simple MLP with one hidden layer for $T_4$. For comparison, we also evaluate the performance of (i) \textit{NP}: randomly initialized vectors as user features, and (ii) \textit{SSP}: behavior latent vectors pretrained on self-supervised $T_1$ as user features. The results in Figure~\ref{fig:case study} indicates that the universal user representations pretrained by PCL achieves better accuracy with faster convergence speed.

\section{Conclusion}
In this paper, we proposed a framework called PCL, which leverages prompts for task adaptation and
knowledge retention in continual user modeling in recommender systems design. 
We proposed position-wise prompts to inherit knowledge from observed tasks and preserve the knowledge learned from each downstream task. 
We also introduced contextual prompts to guide prompt tuning with relations between tasks. 
Our experiments demonstrated the benefits of prompt tuning in continual user modeling and the effectiveness of PCL in various scenarios.

\bibliographystyle{ACM-Reference-Format}
\balance
\bibliography{PromptCL}

\end{document}